\documentstyle[12pt]{article}

\makeatletter
\@addtoreset{equation}{section}

\makeatother

\def\sa{\phantom{a}}
\def\del{\delta}
\def\ord{{\cal O}}

\begin{document}
\begin{flushright}
UT-925
\\
February, 2001
\end{flushright}
\vskip 1.0 truecm

\begin{center}
{\large{\bf  Path integral derivation of
the Brown-Henneaux central charge}}
\end{center}
\vskip .5 cm
\centerline{\bf Hiroaki Terashima}
\vskip .4 cm
\centerline {\it Department of Physics,University of Tokyo}
\centerline {\it Bunkyo-ku,Tokyo 113-0033,Japan}
\vskip 0.5 cm

\begin{abstract}
We rederive the Brown-Henneaux commutation relation and
central charge in the framework of the path integral.
To obtain the Ward-Takahashi identity,
we can use either the asymptotic symmetry or
its leading part.
If we use the asymptotic symmetry,
the central charge arises from the transformation
law of the charge itself.
Thus, this central charge is clearly different from
the quantum anomaly which can be understood as 
the Jacobian factor of the path integral measure.
Alternatively, if we use the leading transformation,
the central charge arises from the fact that
the boundary condition of the path integral is not
invariant under the transformation.
This is in contrast to the usual quantum central charge
which arises from the fact that
the measure of the path integral is not
invariant under the relevant transformation.
Moreover, we discuss the implications of our analysis in
relation to the black hole entropy.
\end{abstract}

\section{Introduction}
To the black hole in general relativity,
we can assign an entropy
called the Bekenstein-Hawking entropy.
There have been many attempts
to understand the origin of this black hole entropy.
In particular, there are some attempts which are based
on the central charge in conformal field theory.
Note that the density of states can be calculated
from the central charge by using the
Cardy formula~\cite{Cardy86,Carlip98}
in conformal field theory.
For example,
in the framework of superstring theory,
it has been shown that
the black hole entropy can be calculated from
the central charge of
the effective superconformal theory
for D-branes~\cite{StrVaf96,CalMal96}.

On the other hand,
Strominger~\cite{Stromi98} has used
the Brown-Henneaux central charge instead of
usual central charges.
In the case of the (2+1) dimensional gravity
with a negative cosmological constant $\Lambda=-1/l^2$,
Brown and Henneaux~\cite{BroHen86} have shown that
the asymptotic symmetry of an asymptotically $AdS_3$
spacetime is the conformal group in 2 dimensions
rather than the $AdS_3$ group, $SO(2,2)$.
Moreover, they have shown that
this symmetry is canonically realized by
the Poisson bracket algebra of the generators
(or the Dirac bracket algebra of the charges)
with a central extension.
The central charge becomes
\begin{equation}
   c=\frac{3l}{2G}.
\end{equation}
By combining this central charge with the Cardy formula,
Strominger has shown that the resultant entropy
agrees with the Bekenstein-Hawking entropy
of the BTZ black hole~\cite{BaTeZa92,BHTZ93}.
Nevertheless, there remain some open questions
in this approach~\cite{Carlip98}.
In particular, the physical meaning of
the Brown-Henneaux central charge is not clear
from the original derivation.

Brown-Henneaux's central charge
was also obtained by some approaches.
Ba\~nados and
co-workers~\cite{Banado95,BaBrOr99} used
the Chern-Simons formulation
of the (2+1) dimensional gravity.
See Refs.~\cite{HenSke98,HyKiLe99,BalKra99}
in the context of AdS/CFT
correspondence~\cite{GuKlPo98,Witten98}.

Natsuume, Okamura and Sato~\cite{NaOkSa00}
have generalized the Brown-Henneaux central charge
to include a conformal scalar field
and applied Strominger's approach to
the MZ black hole~\cite{MarZan96}.
However, since they have obtained
the same charge and central charge as the case of the pure gravity,
the density of states from the Cardy formula does not
agree with the Bekenstein-Hawking entropy.
(The functional form does agree but the over-all numerical
factor does not.)
Thus, they have considered that the Cardy formula gives
the ``maximum possible entropy'' for a given mass.
See also Ref.~\cite{NatOka00} for the extension to various
theories.

Since the Brown-Henneaux central charge is based on
the asymptotic symmetry at infinity in (2+1) dimensions,
Carlip~\cite{Carlip99,Carlip99c} has generalized
to the symmetry near the black hole horizon
in any dimensions.
This symmetry contains a natural Virasoro subalgebra and
the resulting central charge reproduces
the Bekenstein-Hawking entropy by the Cardy formula.
(However, since the generator of Ref.~\cite{Carlip99} is
not ``differentiable''~\cite{ParHo99,Carlip99b},
it has been proposed by Soloviev~\cite{Solovi99} to
use a modified Poisson bracket
which is applicable to the ``non-differentiable''
functionals.)
Similar conclusion was obtained
by Solodukhin~\cite{Solodu99} with
effective 2-dimensional theories
for the spherically symmetric metrics in higher dimensions.
Moreover, Carlip~\cite{Carlip00} has calculated
the prefactor of the Cardy formula
and the logarithmic correction
to the Bekenstein-Hawking entropy.

In this paper, we rederive the Brown-Henneaux
commutation relation and central charge
in terms of the path integral formulation since
it was originally obtained by the canonical formulation.
In view of the equivalence of these two approaches
to quantum theory, we must obtain the same result
within the path integral.
The anomalous commutators have been obtained
also by using the path integral formulation
in various gauge theories~\cite{Fujika86,Fujika87,FujOku98}.
The central charge arises from the quantum anomaly
which is understood as the Jacobian factor of
the path integral measure in usual case~\cite{Fujika87}.
However, the Brown-Henneaux central charge is classical one
because it exists at the level of the Poisson bracket.
We thus want to clarify the origin of this central charge
in the path integral.
Moreover, we want to answer the question
if the Brown-Henneaux central charge is concerned with
the degrees of freedom contained in the system
as usual central charges.
We also hope that
this path integral derivation would be useful
for exploring the relation between
Strominger's approach
and Gibbons-Hawking's approach~\cite{GibHaw77,Hawkin79}.
This is because the topological consideration
is an advantage of the path integral calculation.

One might think that the path integral
would be irrelevant to the classical quantity.
However, the classical quantity still exists after
the quantization as the zeroth order term in $\hbar$.
In order to calculate the whole (classical and quantum)
central charge by the path integral,
we need the derivation of its classical part
within the framework of the path integral.
Although the quantum part is difficult
to calculate,
its origin in terms of the path integral
is quite clear.
We thus ignore the quantum part
and concentrate on the classical one.

To derive the commutation relation,
we must first identify the charge.
Since the original derivation was based on
the Regge-Teitelboim method~\cite{RegTei74},
the charge was obtained indirectly
by integrating its variation.
This process seems to be ad hoc in more general
situations.
Thus, we extract the charge from the variation of
the action, directly.
We then use two transformations
to obtain the Ward-Takahashi identity.
One is the asymptotic symmetry and the other is
its leading part.
If we use the asymptotic symmetry,
we find that 
the central charge arises from
the transformation law of the charge itself.
Thus, we can see it as a classical central charge.
On the other hand,
if we use its leading transformation,
we find that the central charge arises due to
the change of the {\em boundary condition} of
the path integral.
This contrasts with the usual quantum central charge
which arises due to the change of the {\em measure} of
the path integral.

The paper is organized as follows.
In Sec.~\ref{action}, we summarize the variation
and transformation property of the action.
In Sec.~\ref{ads}, we review the asymptotically $AdS_3$
spacetime and calculate some basic quantities.
In Sec.~\ref{charge},
we identify the Brown-Henneaux charge
and derive its transformation law.
In Sec.~\ref{commutator},
we combine all the results to obtain the commutation
relation.
In Sec.~\ref{another},
we give another derivation where the origin of
the central charge is more interesting.
In Sec.~\ref{discuss},
we discuss the implications of our result.

\section{Variation of Action}
\label{action}
We consider the (2+1)-dimensional gravity
with a negative cosmological constant $\Lambda<0$.
We assume that the boundary of the spacetime is only
at infinity $\Sigma^\infty$
whose unit normal vector is $u^a$.
We thus begin with the Hilbert action with
the surface term~\cite{GibHaw77,Hawkin79},
\begin{equation}
 S = \frac{1}{16\pi G}\int_M \sqrt{-g}\;\left(R-2\Lambda\right) \;d^3x+
   \frac{1}{8\pi G}\int_{\Sigma^\infty}\sqrt{-\gamma}\;\Theta \;d^2x,
\end{equation}
where $\gamma_{ab}$ is the induced metric on
$\Sigma^\infty$ defined by
\begin{equation}
  \gamma_{ab}\equiv g_{ab}-u_a u_b,
\end{equation}
and $\Theta^{ab}$ is the extrinsic curvature of $\Sigma^\infty$
defined by
\begin{equation}
\Theta^{ab}=\gamma^{ac}\nabla_c u^b, \qquad
\Theta=g_{ab} \Theta^{ab}=\gamma^{ab}\nabla_a u_b.
\end{equation}

As is well known, the variation of this action with the condition
$\del g_{ab}=0$ at $\Sigma^\infty$ contains
no surface terms and thus we can obtain Einstein equation.
However, if we take a {\em generic} variation of this action,
we obtain that~\cite{York86,BroYor93}
\begin{equation}
 \del S=-\frac{1}{16\pi G}\int_M \sqrt{-g}\;\tilde{G}^{ab}\;\del g_{ab}\;d^3x
  -\frac{1}{16\pi G}\int_{\Sigma^\infty}\sqrt{-\gamma}\;
     \Pi^{ab}\;\del\gamma_{ab}\;d^2x,
\label{delS}
\end{equation}
where
\begin{eqnarray}
  \tilde{G}^{ab}&=&R^{ab}-\frac{1}{2}g^{ab}R+\Lambda g^{ab}, \\
  \Pi^{ab}  &=& \Theta^{ab} - \Theta\gamma^{ab}.
\end{eqnarray}
(Note that $\sqrt{-\gamma}\;\Pi^{ab}$ has the same form as
the canonical momentum in ADM formulation.)

Next, we consider the coordinate transformation
which is generated by a vector $\zeta^a$, namely,
\begin{equation}
   \del_{\zeta} g_{ab} = \pounds_\zeta g_{ab} 
               = \nabla_a\zeta_b+\nabla_b\zeta_a.
\end{equation}
By using Eq.~(\ref{delS}), we find that
\begin{equation}
 \del_{\zeta} S = 
  -\frac{1}{8\pi G}\int_M \sqrt{-g}\;\tilde{G}^{ab}\;
     \nabla_a\zeta_b\;d^3x
  -\frac{1}{8\pi G}\int_{\Sigma^\infty}\sqrt{-\gamma}\;
     \Pi^{ab}\;\nabla_a\zeta_b\;d^2x,
\end{equation}
where we have used
\begin{equation}
  \Pi^{ab}\del\gamma_{ab}=\Pi^{ab}\del g_{ab},
\end{equation}
since $\Pi^{ab}u_a=0$.
We then decompose $\zeta^a$ into
$\tilde{\zeta}^a$ and $\hat{\zeta}^a$, where
\begin{eqnarray}
  \tilde{\zeta}^a &\equiv & \gamma^a_{\sa b} \zeta^b=
                            \zeta^a - \eta \,u^a, \\
  \hat{\zeta}^a   &\equiv &  \eta \,u^a, \\
  \eta  &  = & \zeta^a u_a.
\end{eqnarray}
Note that $\tilde{\zeta}^a$ is tangential to
the boundary $\Sigma^\infty$ since $\tilde{\zeta}^a u_a=0$
and $\hat{\zeta}^a$ is normal to the boundary
because $\hat{\zeta}^a$ is proportional to the normal vector $u^a$.
The tangential part becomes
\begin{equation}
\Pi^{ab}\;\nabla_a \tilde{\zeta}_b=
 \Pi^{cd}\,\gamma_c^{\sa a}\gamma_d^{\sa b}
  \,\nabla_a \tilde{\zeta}_b=\Pi^{cd}\;{\cal D}_c \tilde{\zeta}_d,
\end{equation}
where ${\cal D}_a$ is the covariant derivative associated
with $\gamma_{ab}$.
On the other hand, the normal part becomes
\begin{equation}
\Pi^{ab}\;\nabla_a \hat{\zeta}_b=
 \eta\,\Pi^{ab}\,\nabla_a u_b=
  \eta\,\Pi^{ab}\,\Theta_{ab}=
  \eta\,\left(\Theta^{ab}\,\Theta_{ab}-\Theta^2\right).
\end{equation}
Therefore, one finds that
\begin{eqnarray}
 \del_{\zeta} S &=& 
 -\frac{1}{8\pi G}\int_M \sqrt{-g}\;\tilde{G}^{ab}\;\nabla_a\zeta_b\;d^3x
  \nonumber \\
  & &
{}-\frac{1}{8\pi G}\int_{\Sigma^\infty}\sqrt{-\gamma}\;
  \left[ \Pi^{ab}\;{\cal D}_a \tilde{\zeta}_b +
   \eta\,\left(\Theta^{ab}\,\Theta_{ab}-\Theta^2\right)\right]\;d^2x.
     \label{delxiS}
\end{eqnarray}

\section{Asymptotically $AdS_3$ Spacetime}
\label{ads}
From now on, we consider the asymptotically $AdS_3$
spacetime which is defined by the boundary condition,
\begin{eqnarray}
  g_{tt} &=&-\frac{r^2}{l^2}+\ord(1),         \nonumber \\
  g_{tr} &=& \ord(1/r^3),                     \nonumber \\
  g_{t\phi} &=& \ord(1),                      \label{aads}\\
  g_{rr}  &=& \frac{l^2}{r^2}+\ord(1/r^4),    \nonumber \\
  g_{r\phi} &=&\ord(1/r^3),                   \nonumber \\
  g_{\phi\phi} &=& r^2+\ord(1).               \nonumber
\end{eqnarray}
We treat the boundary $\Sigma^\infty$ as the $r=r_\ast$ surface and
then take the limit $r_\ast\to\infty$.

Brown and Henneaux~\cite{BroHen86} have shown that
the asymptotic symmetry, namely the coordinate transformation
which preserves the asymptotic boundary condition (\ref{aads}),
becomes
\begin{eqnarray}
 \xi^t &=&lT(t,\phi)+\frac{l^3}{r^2}\bar{T}(t,\phi)
           +\ord(1/r^4),                     \nonumber \\
 \xi^r &=&rR(t,\phi)+\frac{l^2}{r}\bar{R}(t,\phi)
           +\ord(1/r^3),              \label{expxi}  \\
 \xi^\phi &=& \Phi(t,\phi)+\frac{l^2}{r^2}\bar{\Phi}(t,\phi)
          +\ord(1/r^4),                      \nonumber
\end{eqnarray}
where they satisfy
\begin{eqnarray}
 l\partial_tT(t,\phi) &=& \partial_\phi \Phi(t,\phi)=-R(t,\phi),\nonumber \\
 l\partial_t\Phi(t,\phi)&=&\partial_\phi T(t,\phi),           \label{asym}\\
 \bar{T}(t,\phi)&=&-\frac{l}{2}\partial_tR(t,\phi),             \nonumber \\
 \bar{\Phi}(t,\phi)&=&\frac{1}{2}\partial_\phi R(t,\phi),       \nonumber
\end{eqnarray}
but $\bar{R}(t,\phi)$ is arbitrary, and
this is the conformal group in 2 dimensions.
This fact can be understood easily by means of
conformal infinity by Penrose~\cite{Penros63,Penros64}.
Briefly,
since the induced metric on the boundary $r=r_\ast\to\infty$
is formally written as
\begin{equation}
   \infty \times \left( -dt^2+l^2d\phi^2 \right),
\end{equation}
the conformal transformation on the $(t,\phi)$ plane
leaves this induced metric ``invariant'',
\begin{equation}
   \infty \times e^{\rho(t,\phi)}
  \left( -dt^2+l^2d\phi^2 \right)
   =\infty \times \left( -dt^2+l^2d\phi^2 \right).
\end{equation}

We write the next leading terms of the metric as
\begin{eqnarray}
g_{tt}&=&-\frac{r^2}{l^2}+e_{tt}+\ord(1/r^2),       \nonumber \\
g_{tr}&=&\frac{l^3}{r^3}\,e_{tr}+\ord(1/r^5),       \nonumber \\
g_{t\phi}&=&e_{t\phi}+\ord(1/r^2),                  \label{expg}\\
g_{rr}&=&\frac{l^2}{r^2}+\frac{l^4}{r^4}\,e_{rr}
                              +\ord(1/r^6),         \nonumber \\
g_{r\phi}&=&\frac{l^3}{r^3}\,e_{r\phi}+\ord(1/r^5), \nonumber \\
g_{\phi\phi}&=&r^2+l^2\,e_{\phi\phi}+\ord(1/r^2),   \nonumber
\end{eqnarray}
where $e_{ab}$ depend only on $(t,\phi)$.
Then, one can obtain the expressions
for some basic quantities,
\begin{eqnarray}
 \Pi^t_{\sa t} &=& -\frac{1}{l}+\frac{l}{r^2}
  \left(\frac{1}{2}e_{rr}+e_{\phi\phi} \right)+\ord(1/r^4), \nonumber \\
 \Pi^t_{\sa \phi} &=& \frac{l}{r^2}\,e_{t\phi}+\ord(1/r^4), \nonumber \\
 \Pi^\phi_{\sa t} &=& -\frac{1}{r^2l}\,e_{t\phi}+\ord(1/r^4),         \\
 \Pi^\phi_{\sa \phi} &=&-\frac{1}{l}+\frac{l}{r^2}
  \left(\frac{1}{2}e_{rr}-e_{tt} \right)+\ord(1/r^4), \nonumber \\
 \Theta^{ab}\Theta_{ab}-\Theta^2
  &=& -\frac{2}{l^2}-\frac{2}{r^2}
    \left(e_{tt}-e_{rr}-e_{\phi\phi}\right)+\ord(1/r^4). \nonumber 
\end{eqnarray}
Note that the transformation law for $e_{ab}$ is
\begin{eqnarray}
 \del_{\xi} e_{rr} &=& lT\partial_t e_{rr}+\Phi\partial_\phi e_{rr}
         -2e_{rr}R-4\bar{R},                           \nonumber \\
 \del_{\xi} e_{\phi\phi} &=& 
      lT\partial_t e_{\phi\phi}+\Phi\partial_\phi e_{\phi\phi}
     + \frac{2}{l} e_{t\phi}\partial_\phi T+
      2e_{\phi\phi}\partial_\phi \Phi+2\bar{R} 
        +2\partial_\phi\bar{\Phi},                     \label{etrans} \\
 \del_{\xi} e_{t\phi} &=& lT\partial_t e_{t\phi}+\Phi\partial_\phi e_{t\phi} 
     + e_{tt}l\partial_\phi T+e_{t\phi}\partial_\phi \Phi
       \nonumber \\
     & &\qquad{}+e_{t\phi}l\partial_t T + l^2e_{\phi\phi}\partial_t \Phi
       -l\partial_\phi\bar{T}+l^2\partial_t\bar{\Phi},   \nonumber
\end{eqnarray}
and so on.

Furthermore, the equations of motion $\tilde{G}_{ab}=0$ say that
\begin{eqnarray}
  \frac{l^2}{2}\partial_t e_{rr}
  +l^2\partial_t e_{\phi\phi} &=& \partial_\phi e_{t\phi}, \nonumber \\
  -\frac{1}{2}\partial_\phi e_{rr}
     +\partial_\phi e_{tt} &=& \partial_t e_{t\phi}  ,    \label{eom} \\
   e_{tt}-e_{rr}-e_{\phi\phi}&=&0.                        \nonumber
\end{eqnarray}
These equations arise from the $tr$-, $r\phi$- and $rr$-components.
The other components become trivial up to this order.

\section{Current and Charge}
\label{charge}
We can obtain the Brown-Henneaux charge~\cite{BroHen86} from
the tensor $\Pi^{ab}$,
\begin{equation}
  J[\xi]=-\frac{1}{8\pi G}\lim_{r_\ast\to\infty}
  \int_{r=r_\ast}d\phi\sqrt{\sigma}
  \left(\Pi^a_{\sa b}-\hat{\Pi}^a_{\sa b}\right)\tilde{\xi}^bn_a,
\end{equation}
where $n^a$ is the unit normal vector of the time slice and
$\sigma_{ab}$ is the induced metric on the boundary $r=r_\ast$
of the time slice.
The hat means that it is evaluated by $AdS_3$ spacetime
so that the charge becomes zero in $AdS_3$ spacetime.
This is similar to the charge defined
by Brown and York~\cite{BroYor93} in the context of
the quasilocal energy,
\begin{equation}
  J_{BY}[\xi]=-\frac{1}{8\pi G}
  \lim_{r_\ast\to\infty}
  \int_{r=r_\ast}d\phi\sqrt{\sigma}
  \left(\Pi^{ab}-\hat{\Pi}^{ab}\right)\tilde{\xi}_bn_a.
\end{equation}
However, these are different because
$\hat{\Pi}^a_{\sa b}\tilde{\xi}^b \neq\hat{\Pi}^{ab}\tilde{\xi}_b$
in the subtraction term.

By using the expansions (\ref{expg}) and (\ref{expxi}),
this becomes
\begin{eqnarray}
 J[\xi] &=&\frac{1}{8\pi G}\lim_{r_\ast\to\infty}
     \frac{r_\ast^2}{l}\int_{r=r_\ast}d\phi
  \left(\Pi^t_{\sa b}-\hat{\Pi}^t_{\sa b}\right)\tilde{\xi}^b
  \nonumber \\
   &=&\frac{1}{8\pi G}\int d\phi
  \left\{\left[\frac{1}{2}\left(e_{rr}+1\right)+e_{\phi\phi}\right]lT
        +e_{t\phi}\Phi\right\},
\end{eqnarray}
for the asymptotic symmetry
of the asymptotically $AdS_3$ spacetime.
In fact, one can check that, by using the same expansions,
the Brown-Henneaux charge also becomes this expression.

The current for the transformation is considered as
\begin{equation}
  j^a[\xi]=-\frac{1}{8\pi G}
  \left(\Pi^a_{\sa b}-\hat{\Pi}^a_{\sa b}\right)\tilde{\xi}^b
   \Big|_{r=r_\ast},
\end{equation}
and the charge is then written as
\begin{equation}
  J[\xi]=\lim_{r_\ast\to\infty}
  \int_{r=r_\ast}d\phi\sqrt{\sigma}\, j^a[\xi]\, n_a.
\end{equation}

By using equations of motion (\ref{eom}),
one can find that the charge is actually conserved,
\begin{equation}
  \partial_t J[\xi]=0,
\end{equation}
where we have used the condition (\ref{asym}) for
$\xi^a$.

The charge for the Lie bracket of two vectors becomes
\begin{eqnarray}
 J\Bigl[[\xi_1,\xi_2]\Bigr] &=& \frac{1}{8\pi G}\int d\phi
  \Biggl\{\left[\frac{1}{2}\left(e_{rr}+1\right)
    +e_{\phi\phi}\right]  \nonumber \\
    & & {}\times\left(l^2T_1\partial_t T_2
       +l\Phi_1\partial_\phi T_2-l^2T_2\partial_t T_1
       -l\Phi_2\partial_\phi T_1\right)  \nonumber \\
    & & {}+e_{t\phi}\left(lT_1\partial_t \Phi_2
       +\Phi_1\partial_\phi \Phi_2-lT_2\partial_t \Phi_1
       -\Phi_2\partial_\phi \Phi_1\right)\Biggr\}.
\end{eqnarray}
On the other hand,
by using Eqs. (\ref{etrans}) and (\ref{asym}),
one can obtain that
\begin{eqnarray}
 \del_{\xi_2} J[\xi_1] &=& \frac{1}{8\pi G}\int d\phi
  \left\{\left[\frac{1}{2}\left(\del_{\xi_2} e_{rr}\right)+
       \left(\del_{\xi_2}e_{\phi\phi}\right)\right]lT_1
        +\left(\del_{\xi_2}e_{t\phi}\right)\Phi_1\right\}  \nonumber \\
   &=& J\Bigl[[\xi_1,\xi_2]\Bigr] + K[\xi_1,\xi_2]+\cdots,
   \label{delJ}
\end{eqnarray}
where $K[\xi_1,\xi_2]$ is the Brown-Henneaux
central charge~\cite{BroHen86},
\begin{eqnarray}
  K[\xi_1,\xi_2]&=&\Biggl(J[\xi_1]\quad
  \mbox{at~~$g_{ab}=\hat{g}_{ab}+\pounds_{\xi_2}\hat{g}_{ab}$}
  \Biggr)    \nonumber \\
    &=& -\frac{1}{8\pi G}\int d\phi
  \left[T_1\left(\partial_\phi+\partial_\phi^3\right)
     +\Phi_1\left(l\partial_t+l^3\partial_t^3\right)
   \right]l \Phi_2,
\end{eqnarray}
and ``$\cdots$'' means the terms which vanish
by using the equations of motion (\ref{eom}).
Note that there are the central term in
the transformation law of the charge itself,
which is explicitly derived from
our definition of the charge
without using the Dirac bracket algebra.
However, we must supply the commutator by
using the path integral
in order to identify it as the central charge.
This is because other contributions
might arise from somewhere.
Indeed, in Sec.~\ref{another},
we will see that there is another
possible source of the
classical central charge.

\section{Commutation Relation}
\label{commutator}
To derive the commutation relation of two charges,
we begin with the path integral,
\begin{equation}
\left\langle J[\xi_1] \right\rangle=\int_Bd\mu\,J[\xi_1]\,e^{iS},
\label{pif}
\end{equation}
where $d\mu$ and $B$ denote the measure and boundary condition
of the path integral, respectively.
We first replace the integration variable $g_{ab}$
everywhere in Eq.~(\ref{pif}) with $\tilde{g}_{ab}$.
This step is mathematically trivial,
similar to the replacement
\begin{equation}
 \int\,f(x)\,dx\longrightarrow\int\,f(y)\, dy.
\end{equation}
We recognize this new integration variable as
the transformed metric by the infinitesimal
asymptotic symmetry transformation $\xi_2$,
\begin{equation}
 \tilde{g}_{ab}\equiv g_{ab}+\del_{\xi_2}g_{ab}.
\end{equation}
We then find that
\begin{eqnarray}
 \left\langle J[\xi_1] \right\rangle &=&
    \int_{\tilde{B}} d\tilde{\mu}\, \tilde{J}[\xi_1]\,e^{i\tilde{S}}
   \nonumber \\
  &=& \int_B d\mu\, \left(J[\xi_1]+\del_{\xi_2}J[\xi_1]\right)
     \left(1+i\del_{\xi_2}S \right)  \,e^{iS} \nonumber \\
  &=& \int_B d\mu\, \left(J[\xi_1]+\del_{\xi_2}J[\xi_1]
     +iJ[\xi_1]\del_{\xi_2}S \right)  \,e^{iS} \nonumber \\
 &=& \left\langle J[\xi_1]\right\rangle+
     \left\langle \del_{\xi_2}J[\xi_1]\right\rangle
     +i\left\langle{\rm T}^\ast\,
    J[\xi_1]\del_{\xi_2}S \right\rangle,
\label{WTderi}
\end{eqnarray}
where we have used the fact that the boundary condition
is invariant under the asymptotic symmetry, $\tilde{B}=B$.
We also assumed that the measure of the path integral
is invariant under this transformation, $d\tilde{\mu}=d\mu$,
since we want to see the classical central charge.
Therefore, we can obtain the Ward-Takahashi identity
\begin{equation}
  \left\langle \del_{\xi_2} J[\xi_1] \right\rangle
   =-i \left\langle {\rm T}^\ast\,  J[\xi_1]\; \del_{\xi_2}S 
   \right\rangle.
\label{WT}
\end{equation}

By using Eq.~(\ref{delxiS}), we can evaluate the right-hand side
of this identity.
However, the result becomes infinite
in the limit of $r=r_\ast\to\infty$.
In order to get a finite result,
it is usual to subtract a functional $S_0$ of the
boundary data $\gamma_{ab}$ from the action.
In this case, we choose so that~\cite{BroYor93}
\begin{equation}
 \del_\xi S_0 =
  -\frac{1}{8\pi G}\int_{\Sigma^\infty}\sqrt{-\gamma}\;
  \left[ \hat{\Pi}^a_{\sa b}\;{\cal D}_a \tilde{\xi}^b +
   \eta\,\left(\hat{\Theta}^{ab}\,
     \hat{\Theta}_{ab}-\hat{\Theta}^2\right)\right]\;d^2x,
\end{equation}
where the hats again mean that they are evaluated
by $AdS_3$ spacetime.
For notational simplicity,
we write $S-S_0$ as $S$ anew.

Then, one finds that
\begin{eqnarray}
   \del_{\xi_2}S  &=&  -\frac{1}{8\pi G}\lim_{r_\ast\to\infty}
  \int_{r=r_\ast}dt d\phi\;
   \sqrt{-\gamma}\;\left(\Pi^a_{\sa b}-\hat{\Pi}^a_{\sa b}\right)
   {\cal D}_a \tilde{\xi}_2^b+\cdots    \nonumber \\
     &=&  -\frac{1}{8\pi G}\lim_{r_\ast\to\infty}
    \frac{r_\ast^2}{l}\int_{r=r_\ast} dt d\phi\;
  {\cal D}_a\left[ \left(\Pi^a_{\sa b}-\hat{\Pi}^a_{\sa b}\right)
    \tilde{\xi}_2^b\right]+\cdots    \nonumber \\
    &=&  -\frac{1}{8\pi G}\lim_{r_\ast\to\infty}
    \frac{r_\ast^2}{l}\int_{r=r_\ast} dt d\phi\;
  \partial_t \left[ \left(\Pi^t_{\sa b}-\hat{\Pi}^t_{\sa b}\right)
    \tilde{\xi}_2^b\right]+\cdots    \nonumber \\
     &=& -\int dt \;\partial_t J[\xi_2] +\cdots,
\end{eqnarray}
where ``$\cdots$'' again means the terms which vanish
by using the equations of motion (\ref{eom}).
Note that
\begin{eqnarray}
  \left(\Theta^{ab}\Theta_{ab}-\Theta^2\right)
   -\left(\hat{\Theta}^{ab}\,
 \hat{\Theta}_{ab}-\hat{\Theta}^2\right) &=& \ord(1/r^4), \nonumber \\
 {\cal D}_a\left(\Pi^a_{\sa b}-\hat{\Pi}^a_{\sa b}\right)
    &=& \ord(1/r^4), \nonumber
\end{eqnarray}
by the equations of motion.

Then, the right-hand side of the identity (\ref{WT}) is
\begin{eqnarray}
  \left\langle {\rm T}^\ast\,  J[\xi_1]\; \del_{\xi_2}S  \right\rangle
    &=& -\left\langle {\rm T}^\ast\,  J[\xi_1]\;
           \int dt \;\partial_t J[\xi_2] +\cdots   \right\rangle \nonumber \\
    &=& -\int dt \;\partial_t \left\langle {\rm T}^\ast\,  J[\xi_1]\;
            J[\xi_2] +\cdots   \right\rangle \nonumber \\
    &=& -\int dt \;\partial_t \left\langle {\rm T}\,  J[\xi_1]\;
            J[\xi_2] +\cdots   \right\rangle \nonumber \\
    &=& \left\langle \Bigl[ J[\xi_1],J[\xi_2 ] \Bigr]+\cdots\right\rangle,
\end{eqnarray}
where we have simply identified the T$^\ast$-product
as the T-product
after extracting the time derivative
by the standard Bjorken-Johnson-Low
argument~\cite{Bjorke66,JohLow66}.

Finally, by using Eq.~(\ref{delJ})
and the equations of motion (\ref{eom}),
we can obtain the commutation relation,
\begin{equation}
  \left\langle \Bigl[ J[\xi_1],J[\xi_2 ] \Bigr] \right\rangle= 
   \left\langle i J\Bigl[[\xi_1,\xi_2]\Bigr]
   +i  K[\xi_1,\xi_2]\right\rangle.
\label{BHc}
\end{equation}
This is consistent with
the result of Brown and Henneaux~\cite{BroHen86}
if we identify the Dirac bracket as the commutator,
\begin{equation}
  \{A,B\}_{{\rm D.B.}}
  \longleftrightarrow
   \frac{1}{i}[A,B].
\end{equation}
Note that the classical central charge comes
from the transformation law of the charge (\ref{delJ})
rather than the Jacobian factor of the path integral measure.
This is analogous to
the central charge in $N=2$ supersymmetric
theory~\cite{FujOku98}.

\section{Another Derivation}
\label{another}
We can derive this central charge
in another way~\cite{Terash01}
if we use only the leading part of
the asymptotic symmetry,
\begin{eqnarray}
 \xi'^t &=&lT(t,\phi),          \nonumber \\
 \xi'^r &=&rR(t,\phi),          \label{lead} \\
 \xi'^\phi &=& \Phi(t,\phi),    \nonumber
\end{eqnarray}
where $T,R,\Phi$ again satisfy the condition (\ref{asym}).
This transformation is {\em not} the asymptotic symmetry
since it breaks the boundary conditions
for $g_{tr}$ and $g_{r\phi}$,
\begin{eqnarray}
\pounds_{\xi}g_{tr}  &=& \frac{l^2}{r}
     \partial_t R+\ord(1/r^3),\nonumber \\
\pounds_{\xi}g_{r\phi}  &=& \frac{l^2}{r}
     \partial_\phi R+ \ord(1/r^3).\nonumber
\end{eqnarray}

Note that
\begin{equation}
J[\xi']=J[\xi],  \qquad
J\Bigl[[\xi'_1,\xi'_2]\Bigr]=J\Bigl[[\xi_1,\xi_2]\Bigr],
\end{equation}
and
\begin{equation}
\del_{\xi'} S=\del_{\xi} S,
\end{equation}
since the non-leading parts do not contribute to these
quantities.
On the other hand, we have
\begin{equation}
 \del_{\xi'_2} J[\xi'_1]=
   J\Bigl[[\xi'_1,\xi'_2]\Bigr] + K'[\xi'_1,\xi'_2]+\cdots,
\label{delJ2}
\end{equation}
where
\begin{equation}
 K'[\xi'_1,\xi'_2] = -\frac{1}{8\pi G}\int d\phi
  \left[T_1\partial_\phi+\Phi_1 l\partial_t\right]l \Phi_2.
\end{equation}
This is not a non-trivial central charge since
we can eliminate this term by adding a constant to the charge.
(Actually, it can be achieved by choosing
the $M=J=0$ BTZ black hole as the background rather than
$AdS_3$ spacetime~\cite{Terash01}.)
This is the interesting aspect of this
leading transformation (\ref{lead}).
Since the remaining quantities in the
Ward-Takahashi identity (\ref{WT}) are equal,
one might think that one could obtain
the commutator without the non-trivial central charge,
\begin{equation}
  \left\langle \Bigl[ J[\xi_1],J[\xi_2 ] \Bigr] \right\rangle
   \stackrel{??}{=}
   \left\langle i J\Bigl[[\xi_1,\xi_2]\Bigr]
   +i  K'[\xi'_1,\xi'_2]\right\rangle,
\end{equation}
if we use this leading transformation (\ref{lead}).
However, this is not correct.
Since the leading transformation breaks the boundary condition
of the path integral $B$, we cannot obtain
the Ward-Takahashi identity (\ref{WT}).
Instead, the Ward-Takahashi identity is supplemented with
an additional term due to
the change of the boundary condition $B$, namely,
\begin{equation}
 \left\langle \del_{\xi'_2} J[\xi'_1] \right\rangle
 =-i \left\langle {\rm T}^\ast\,J[\xi'_1]\;\del_{\xi'_2}S\right\rangle
  -\Delta[\xi'_1,\xi'_2],
\end{equation}
where
\begin{equation}
 \Delta[\xi'_1,\xi'_2] \equiv 
   \left( \int_{B+\del_{\xi'_2}B} -
      \int_B\right) d\mu\, J[\xi'_1] \, e^{iS},
\end{equation}
and the boundary condition $B+\del_{\xi'_2}B$ denotes
that the transformed metric $g_{ab}+\del_{\xi'_2}g_{ab}$ must
satisfy the asymptotically $AdS_3$ condition (\ref{aads}).
Repeating the derivation as above,
we find that
\begin{equation}
  \left\langle \Bigl[ J[\xi'_1],J[\xi'_2 ] \Bigr] \right\rangle= 
   \left\langle i J\Bigl[[\xi'_1,\xi'_2]\Bigr]
   +i  K'[\xi'_1,\xi'_2]\right\rangle+i\Delta[\xi'_1,\xi'_2].
\end{equation}

In order to evaluate $\Delta[\xi'_1,\xi'_2]$,
we perform the infinitesimal change of
the integration variable corresponding to
the inverse transformation of $\xi_2'$ in the first integral
and that of $\xi_2$ in the second integral,
similar to the calculations in Eq.~(\ref{WTderi}).
The integrals then become
\begin{eqnarray}
  \int_{B+\del_{\xi'_2}B}d\mu\, J[\xi'_1] \, e^{iS}
   &=& \int_B d\mu\left(J[\xi'_1]-iJ[\xi'_1]\,\del_{\xi'_2}S
                -\del_{\xi'_2} J[\xi'_1] \right) \, e^{iS}, \\
  \int_B d\mu\,J[\xi'_1] \, e^{iS}
   &=& \int_B d\mu\left(J[\xi'_1]-iJ[\xi'_1]\,\del_{\xi_2}S
                -\del_{\xi_2} J[\xi'_1] \right) \, e^{iS}.
\end{eqnarray}
Note that the boundary condition of both of the path integral
become the same.
By using Eqs.~(\ref{delJ}) and (\ref{delJ2}),
we can obtain that
\begin{eqnarray}
  \Delta[\xi'_1,\xi'_2] &=& \left\langle 
      \del_{\xi_2} J[\xi'_1]-\del_{\xi'_2} J[\xi'_1]
       \right\rangle \nonumber \\
        &=& \left\langle 
      K[\xi_1,\xi_2]- K'[\xi'_1,\xi'_2] \right\rangle.
\end{eqnarray}
Thus, we can again obtain the Brown-Henneaux commutation relation,
\begin{equation}
   \left\langle \Bigl[ J[\xi'_1],J[\xi'_2 ] \Bigr] \right\rangle= 
   \left\langle i J\Bigl[[\xi'_1,\xi'_2]\Bigr]+
     iK[\xi_1,\xi_2]\right\rangle,
\end{equation}
by using the equations of motion.

Therefore, the non-trivial part of the central charge
arises from the fact that the {\em boundary condition} of
the path integral $B$
is not invariant under the leading transformation $\xi'$.
This phenomenon is in contrast to the usual quantum case
where the anomaly arises from the fact that the {\em measure} of
the path integral $d\mu$
is not invariant under the relevant transformation.

\section{Discussion}
\label{discuss}
We have reproduced the Brown-Henneaux commutation relation
in the context of the path integral.
The origin of the central charge is not the Jacobian factor
of the path integral measure as the quantum anomaly.
If we use the asymptotic symmetry to derive the Ward-Takahashi
identity, it arises from the transformation law of the charge.
Thus, it can be considered as a classical anomaly
even though the path integral formulation
has been used to obtain the commutator of two charges.
This is similar to
the central charge in $N=2$ supersymmetric
theory.
In order to apply to the black hole entropy,
we want to relate this central charge with
some ignorance.
It would be considered as follows.
We could see the tensor $\Pi^{ab}$ as a tensor
in 2 dimensions on the boundary $\Sigma^\infty$ since
it can be obtained as the conjugate variable
to the induced metric $\gamma_{ab}$ on the boundary.
The charge $J[\xi]$ is made from this tensor and
boundary value of $\xi$.
In this sense, the charge is a quantity on the 2-dimensional
boundary.
However, the transformation $\xi$ is
in the 3-dimensional spacetime.
This transformation consists of the
2-dimensional transformation with some additional terms.
Especially, $\bar{T}$ and $\bar{\Phi}$ terms,
which are required to maintain the boundary condition,
give rise to the central charge.
This gap corresponds to the ignorance due to
the limit $r_\ast\to\infty$.

Alternatively, in order to derive the Ward-Takahashi
identity, we can use the leading part of the
asymptotic symmetry.
Then, we can find that the central charge arises from
the fact that the {\em boundary condition}
of the path integral
is not invariant under the leading transformation.
This is in contrast to the quantum central charge
which arises from the fact that
the {\em measure} of the path integral is not
invariant under the relevant transformation.
From this picture, we could see that
the Brown-Henneaux central charge does not count 
the degrees of freedom in the system.
Moreover, we could understand
other classical central charges,
such as in $N=2$ supersymmetric theory,
as above in the path integral formulation.
Our analysis also suggests the possibility that the
classical central charge may arise in more general theories
if the boundary condition of the path integral is non-trivial.

Finally, one of the advantage of the present analysis is that
the charge is identified directly
and it is thus easy to apply to
more general situations.
The past approaches were based
on the Regge-Teitelboim method
where the charge $J[\xi]$ was derived from
its variation $\del J[\xi]$.
However, in more general cases~\cite{Carlip99,NaOkSa00},
it is not straightforward to do this integration.
On the other hand,
since the present analysis can identify
the charge directly,
it would be straightforward to apply
in such cases.
The other advantage of the present analysis
would be the topological consideration.
We hope that
we could relate Strominger's approach
to Gibbons-Hawking's approach~\cite{GibHaw77}
by using this path integral derivation.

\section*{Acknowledgments}
The author thanks K. Fujikawa for introducing him
to this subject and helpful discussions.
The author also thanks Y. Shibusa for comments
and discussions.

%\bibliographystyle{prsty}
%\bibliography{phys}

\end{document}